\documentclass[ reprint,
superscriptaddress,
showpacs,amsfonts,amsmath,amssymb,aps,prb,floatfix
]{revtex4-1}
\usepackage{graphicx}
\usepackage{dcolumn}
\usepackage{bm}
\usepackage{epsfig}
\usepackage{array}
\usepackage{hyperref}
\usepackage{multirow}
\begin{document}
\title{The High Temperature Phase Transitions of Hexagonal \texorpdfstring{YMnO$_{3}$}{YMnO3}}
\author{Alexandra S. Gibbs}
\affiliation{School of Chemistry and EaStCHEM, University of St Andrews, North Haugh, St Andrews KY16 9ST, United Kingdom}
\affiliation{Scottish Universities Physics Alliance (SUPA), School of Physics and Astronomy, University of St Andrews, North Haugh, St Andrews KY16 9SS, United Kingdom}
\author{Kevin S. Knight}
\affiliation{ISIS Facility, Rutherford Appleton Laboratory, Chilton, Didcot OX11 OQX, United Kingdom}
\author{Philip Lightfoot}
\email[To whom correspondence should be addressed: ]{pl@st-andrews.ac.uk}
\affiliation{School of Chemistry and EaStCHEM, University of St Andrews, North Haugh, St Andrews KY16 9ST, United Kingdom}
%\date{\today}
\begin{abstract}
We report a detailed high-resolution powder neutron diffraction investigation of the structural behaviour of the multiferroic hexagonal polymorph of YMnO$_{3}$ between room temperature and 1403~K. The study was aimed at resolving previous uncertainties regarding the nature of the paraelectric-ferroelectric transition and the possibilities of any secondary structural transitions. We observe a clear transition at 1258$\pm$14~K corresponding to a unit cell tripling and a change in space group from centrosymmetric $P6_{3}/mmc$ to polar $P6_{3}cm$. Despite the fact that this symmetry permits ferroelectricity, our experimental data for this transition analysed in terms of symmetry-adapted displacement modes clearly supports previous theoretical analysis that the transition is driven primarily by the antiferrodistortive $K_{3}$ mode. We therefore verify previous suggestions that YMnO$_{3}$ is an improper ferrielectric. Furthermore, our data confirm that the previously suggested intermediate phase with space group $P6_{3}/mcm$ does not occur. However, we do find evidence for an isosymmetric phase transition (i.e. $P6_{3}cm$ to $P6_{3}cm$) at $\approx$920~K which involves a sharp decrease in polarization. This secondary transition correlates well with several previous reports of anomalies in physical properties in this temperature region and may be related to Y-O hybridization.
\end{abstract}
\maketitle
\section{Introduction}
\begin{figure}[b!]
		\includegraphics[width=1.0\columnwidth]{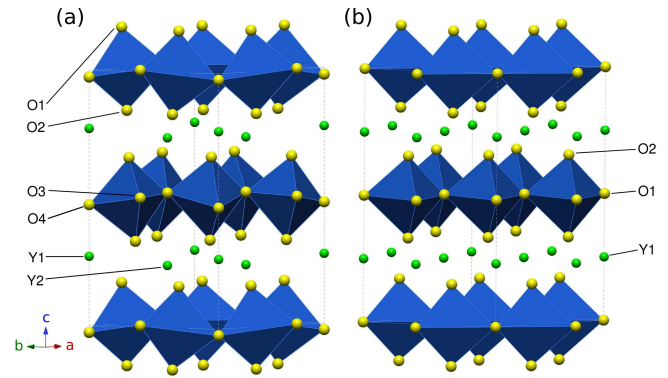}
	\caption{\label{fig:fig1}(Colour Online) (a) The ambient temperature polar $P6_{3}cm$ structure of hexagonal YMnO$_{3}$. (b) The high temperature centrosymmetric $P6_{3}/mmc$ form displayed in the $P6_{3}cm$ basis to allow comparison (transformation matrix:$\left[a+b,-a+2b,c+\frac{1}{4}\right]$). The Y atoms (green) are 8-fold coordinated by oxygen (yellow) and the Mn atoms are 5-fold coordinated (blue polyhedra).}
\end{figure}

The \textit{A}MnO$_{3}$ manganites (with \textit{A}= Lanthanide, In, Y, Sc) have attracted much interest in recent years due to their multiferroic properties \cite{Lee2008,Choi2010}. Two structural forms of these materials exist, both displaying multiferroicity\cite{Zhou2006}. The orthorhombic form, a perovskite with room temperature space group \textit{Pnma}, occurs for \textit{A}=La-Tb. The hexagonal form, with a layered structure, shown in Figure \ref{fig:fig1}(a) with space group $P6_{3}cm$ at room temperature, is favoured for \textit{A}= Dy-Lu, In, Y, Sc. Varying the synthesis technique, however, allows for some flexibility of this trend\cite{Zhou2006}.

YMnO$_{3}$ can be synthesised in either of the two polymorphs. The hexagonal form is obtained when standard ambient pressure solid state synthesis conditions are used. This form has Mn$^{3+}$ ions coordinated by five oxide ions, forming a trigonal bipyramid. The Y$^{3+}$ ions are coordinated by eight oxide ions (six equatorial oxygens from two symmetry inequivalent sites and two inequivalent apical oxygens). The bipyramids are tilted with respect to the \textit{c}-axis and the two Y-O apical bond lengths for each yttrium site are unequal as a result of this.

The hexagonal manganites are ferrielectric (i.e. have opposite but unequal dipole moments within the unit cell leading to a net polarization) up to high temperatures in excess of $T_{C}\!\approx\,$900~K \cite{Abrahams2001} and order antiferromagnetically below $T_{N}\!\!\approx\,$70~K\cite{Chatterji2007}. The ferrielectricity is due to opposing dipoles caused by opposite but unequal displacements of the two yttrium sites and the associated tilting and distortion of the MnO$_{5}$ bipyramids\cite{Fennie2005,VanAken2004}. This is an unusual driving mechanism for ferroelectricity and has been termed `geometric ferroelectricity'\cite{VanAken2004} as it seems to depend purely on ionic size effects rather than the more ubiquitous electronic effects such as $d^{0}$ cation Jahn-Teller distortions (Ti$^{4+}$, Nb$^{5+}$ etc.)\cite{Khomskii2006} or the presence of stereochemically active $s^{2}$ lone pair cations such as Pb$^{2+}$ or Bi$^{3+}$.

Due to the experimental difficulties in measuring physical properties such as the dielectric constant at the high temperatures of the ferrielectric phase transition, ambiguity exists about the mechanism and exact nature of the transition between the room temperature polar $P6_{3}cm$ structure and the centrosymmetric $P6_{3}/mmc$ state (shown in Figure \ref{fig:fig1}(b)) which is suggested to exist above $\approx$1250~K. This aristotype structure is the undistorted form of the ambient temperature structure. The low temperature structure is generated from the aristotype by loss of mirror symmetry perpendicular to the \textit{c}-axis resulting in tilted MnO$_{5}$ bipyramids, unequal apical Y-O bond lengths and a larger unit cell, with three times the $P6_{3}/mmc$ unit cell volume and the low temperature unit cell parameters \textit{a}=$\sqrt{3}a^{\prime}$ and \textit{c}=\textit{c$^{\prime}$}.

Previous studies have produced a wide range of transition temperatures using various methods, tabulated in Table \ref{tab:tab1}, for both the ferrielectric transition temperature ($T_{C}$) and the unit cell tripling transition between $P6_{3}/mmc$ and $P6_{3}cm$ ($T_{S}$) which tilts the MnO$_{5}$ trigonal bipyramids (untilted in the high-temperature $P6_{3}/mmc$ phase) and corrugates the Y$^{3+}$ layers leading to tripling of the \textit{ab}-plane unit cell area and therefore the unit cell volume.

\begin{table}[b!]
\caption{\label{tab:tab1}The previously reported transition temperatures of the ferrielectric ($T_{C}$) and unit cell tripling ($T_{S}$) transitions. PXRD = powder x-ray diffraction, SXD = single crystal x-ray diffraction, PND = powder neutron diffraction, DTC = differential thermal calorimetry, MEM = maximum entropy method}
\begin{ruledtabular}
\begin{tabular}{l c c c }
Reference & \textit{T$_{C}$} (K) & \textit{T$_{S}$} (K) & Method\\\hline
Ismailzade & \multirow{2}{*}{933}& \multirow{2}{*}{-}&pyroelectric\\
and Kizhaev\cite{Ismailzade1965}  & & &current\\
\L ukaszewicz and  &\multirow{2}{*}{-}& \multirow{2}{*}{1275}& \multirow{2}{*}{SXD} \\
Karat-Kalici$\acute{\textrm{n}}$ska \cite{Luka1974} & & & \\
Katsufuji \textit{et al.} (2001) \cite{Katsufuji2001} & 910 & - & resistivity\\
Katsufuji \textit{et al.} (2002)\cite{Katsufuji2002} & $\geq$1000 &$\geq$1000  & PXRD  \\
N$\acute{\textrm{e}}$nert \textit{et al.} (2005)\cite{Nenert2005} & 1020 &1273 & SXD \\
N$\acute{\textrm{e}}$nert \textit{et al.} (2007)\cite{Nenert2007} & 1125 & 1350 & powder DTC\\
Jeong \textit{et al.}\cite{Jeong2007} & -& $\geq$1200& PND\\
Choi \textit{et al.}\cite{Choi2010} & $\approx$880 & - & resistivity\\
Kim \textit{et al.}\cite{Kim2009} & $\approx$920& -&  PXRD (MEM)\\
\end{tabular}
\end{ruledtabular}
\end{table}
\begin{figure}[t!]
		\includegraphics[width=1.0\columnwidth]{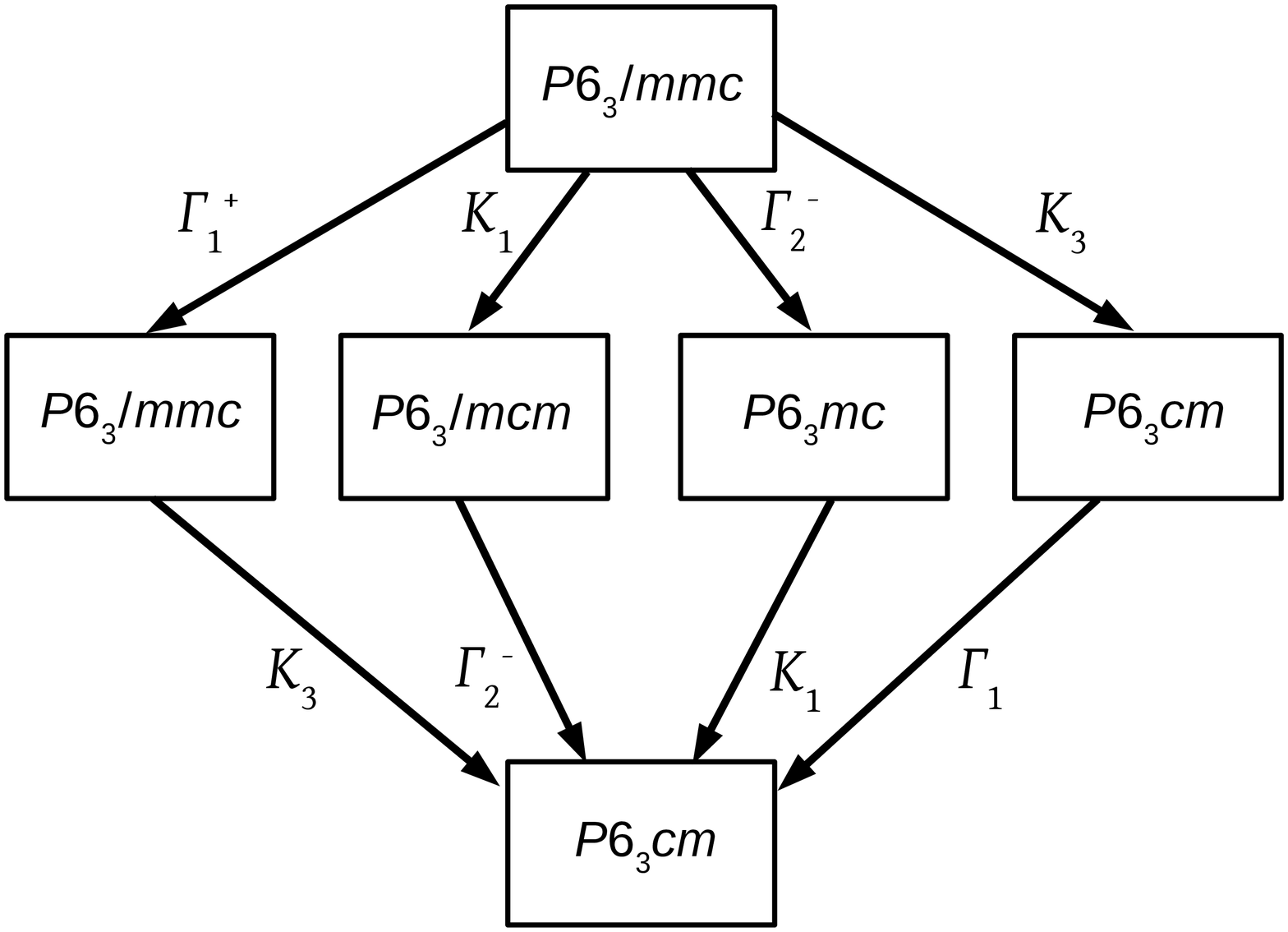}
	\caption{\label{fig:fig2}The descent-of-symmetry diagram for the transition between centrosymmetric $P6_{3}/mmc$ and polar $P6_{3}cm$. The arrow labels denote the distortion modes leading to the lower symmetry space groups, the \textit{K} modes are unit cell tripling modes.}
\end{figure}
The variety of transition temperatures measured has been attributed to impurities introduced by the synthesis method \cite{Nenert2007} but the subtle nature of the transitions and lack of polarization or thermodynamic measurements also make transition temperatures difficult to define.

Furthermore, the possibility of an intermediate phase between $P6_{3}/mmc$ and $P6_{3}cm$ has been noted\cite{Lonkai2004,Nenert2005,Abrahams2009}. The possible intermediate phases were identified using descent of symmetry arguments \cite{Fennie2005,Lonkai2004}; a diagram of the possible transition paths is shown in Figure \ref{fig:fig2}. The transition from $P6_{3}/mmc$ to $P6_{3}/mcm$ would involve unit cell tripling caused by displacement of the O1-Mn-O1 axis within the \textit{ab}-plane but no tilt of the MnO$_{5}$ bipyramids or Y$^{3+}$ displacement. Taking the path from $P6_{3}/mmc$ to $P6_{3}mc$ would involve no change in unit cell volume, only loss of the mirror plane allowing independent polar displacements of all atoms along \textit{c}. Therefore, the most straightforward method of identifying the transition path is to determine whether the cell tripling occurs at the same point as the polar displacement and examine the Y site splitting and tilt of the MnO$_{5}$ bipyramids.

Although N$\acute{\textrm{e}}$nert  \textit{et al.}\cite{Nenert2007} proposed $P6_{3}/mcm$ as the intermediate phase, no structural parameters from their intermediate phase region were published. Therefore, a systematic variable temperature study with careful examination of the distortions of the structure and comparative refinements of alternative space groups is required.

Most crystallographic studies thus far have used powder x-ray diffraction which usually does not allow such reliable and precise determination of lattice parameters and atomic positions in oxide materials as neutron diffraction. This therefore increases the difficulty of pinning down the location and nature of the phase transitions. The only previous powder neutron study \cite{Jeong2007} involved four temperatures between 1000~K and 1400~K and left the key issues unresolved. To attempt to resolve this uncertainty, we have undertaken a higher resolution powder neutron diffraction study using finer temperature intervals.
\section{Experimental}
\subsection{Sample Synthesis and Characterisation}
A single phase polycrystalline sample was prepared by standard solid state synthesis. A stoichiometric mixture of Y$_{2}$O$_{3}$ (Sigma Aldrich 99.999\%) and MnO$_{2}$ (Sigma Aldrich 99.99+\%) was ground under acetone, pressed into pellets and heated at 1473~K on sacrificial powder in an alumina boat for 140 hours with intermediate grindings every 18 hours. The sample quality was monitored using laboratory x-ray diffraction (Stoe STADI P with Cu K$_{\alpha1}$ source in flat-plate transmission mode) throughout the synthesis to ensure a phase-pure sample was obtained. Energy Dispersive X-ray (EDX) spectroscopy was also used to confirm that the sample was not contaminated by, for example, aluminium from the alumina crucible.\subsection{Powder Neutron Diffraction}
Powder neutron diffraction was undertaken on the High Resolution Powder Diffractometer (HRPD) at ISIS\cite{Ibberson1992,Ibberson200947}. A 5g sample was sealed in a quartz tube and placed in a cylindrical vanadium can mounted in a standard furnace. Data were collected at 28 temperatures between 293~K and 1403~K (every 60~K from 373~K to 1093~K then intervals of 10~-~30~K to 1403~K) with appropriate equilibration times at each temperature before commencement of data collection.
The data used for the analysis were all taken from the backscattering detector bank centred at 168$^{\circ}$ with resolution $\frac{\Delta d}{d}\approx4-5\times10^{-4}$. Data were refined using the Rietveld method with the program GSAS \cite{gsasref}. A 20 term shifted Chebyshev background function was used to account for the substantial quartz background and appropriate absorption corrections were applied. Small peaks from the vanadium can were identified in all patterns, these were not included in the refinements.

\section{Results}
\subsection{High Temperature Phase}
\begin{figure}[t!]
		\includegraphics[width=1.0\columnwidth]{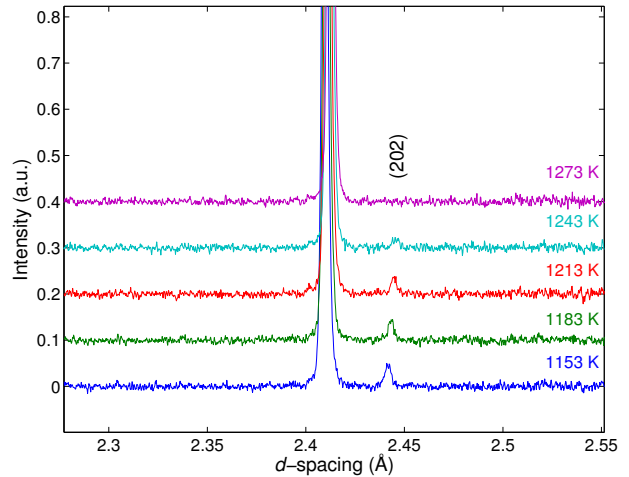}
	\caption{\label{fig:fig3}(Colour Online) A portion of the raw diffraction patterns for 1153~K, 1183~K, 1213~K, 1243~K and 1273~K offset by 0.1 on the intensity axis for clarity. The appearance of the (202) peak at 2.44~\AA\ signals the entry into the polar $P6_{3}cm$ phase.}
\end{figure}

The transition from the high temperature $P6_{3}/mmc$ phase to the low temperature unit cell tripled phase is signalled most clearly by the appearance of the $\left( 202 \right)$ peak due to the factor of $\sqrt{3}$ increase in \textit{a} as shown in Figure \ref{fig:fig3}. This peak is absent in all datasets above the 1243~K dataset where it first appears. It then increases in intensity with decreasing temperature. The diffraction patterns have no reflections breaking $P6_{3}/mmc$ symmetry at 1273~K or above. 
\begin{figure}[t!]
		\includegraphics[width=1.0\columnwidth]{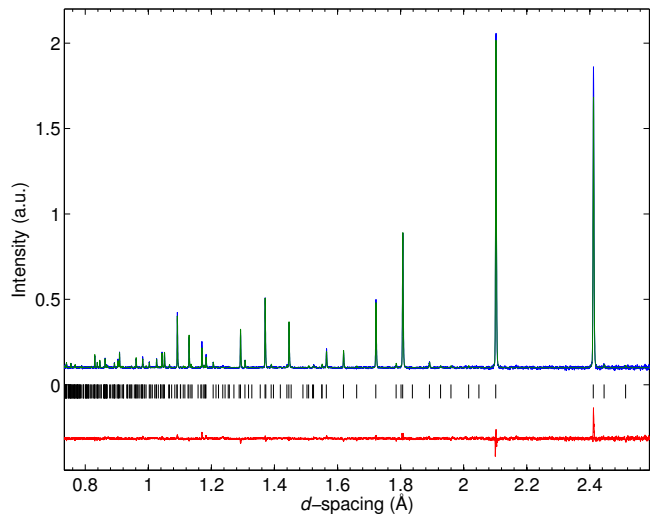}
	\caption{\label{fig:fig4}(Colour Online) The full-range fit resulting from the Rietveld refinement of the 1243~K dataset with the $P6_{3}cm$ model. The blue line is the raw data, the green line the {Rietveld} fit, the red line the difference profile and the black markers indicate predicted reflection positions.}
	\end{figure}
The Rietveld refinements in the high temperature phase converged quickly and anisotropic thermal displacement factors were refined for all sites. Refined structural parameters in the $P6_{3}/mmc$ phase at 1303~K are given in Table \ref{tab:tab2}.

\begin{table*}[t!]
\caption{\label{tab:tab2}The structural parameters for the 1303~K data refined in $P6_{3}/mmc$ (No.~194). The lattice parameters are \textit{a}=3.618961(5)~\AA\ and \textit{c}=11.34090(3)~\AA. The refinement gave \textit{wR}$_{p}$=0.0331, $\chi^{2}$=6.356. The variables $u_{11}$, $u_{22}$, $u_{33}$ and $u_{12}$ represent the anisotropic thermal displacement parameters.}
\begin{ruledtabular}
\begin{tabular}{c c c c c d d d}
\multirow{2}{*}{Atom}&Wyckoff&\multirow{2}{*}{\textit{x}}&\multirow{2}{*}{\textit{y}}&\multirow{2}{*}{\textit{z}}&\multicolumn{1}{c}{\multirow{2}{*}{$u_{11}(=u_{22})\times$ 100 (\AA$^{2}$)}}&\multicolumn{1}{c}{\multirow{2}{*}{$u_{33}\times$ 100 (\AA$^{2}$)}}&\multicolumn{1}{c}{\multirow{2}{*}{$u_{12}\times$ 100 (\AA$^{2}$)}}\\
&Position&&&&&&\\\hline
Y1 & 2a&0 &0&0&2.01(3)&6.75(5)&1.004(16)\\
Mn & 2c&$\frac{1}{3}$&$\frac{2}{3}$&$\frac{1}{4}$&3.97\text{(5)}&1.61(7)&1.99(2)\\
O1 & 2b&0& 0&$\frac{1}{4}$&3.35\text{(4)}&6.64(8)&1.69(2)\\
O2 & 4f&$\frac{1}{3}$&$\frac{2}{3}$&0.08557(7)&3.52\text{(3)}&1.84(4)&1.760(16)\\
\end{tabular}
\end{ruledtabular}
\end{table*}
The smooth variation of the thermal displacement factors (see supplementary material) and lattice parameters with temperature in this high temperature regime suggest that there is no higher temperature phase transition. Refinements in the polar space group $P6_{3}mc$ in the same unit cell did not give improved fits (see supplementary material). Moreover, there is no previous evidence of any physical property measurements supporting the occurrence of a non-centrosymmetric phase at this temperature. There is also no indication of a transition to a further high temperature polymorph in space group $P6/mmm$ which was recently proposed by Abrahams \cite{Abrahams2009} to be the aristotype structure present above the $P6_{3}/mmc$ phase. This structure would have a halved \textit{c}-axis length relative to the other phases and would require a large displacement ($\approx$~0.99~\AA) in oxide ion positions. The existence of this phase also seems implausible in chemical grounds as it requires abnormally short Y-O bonds. From our data we conclude that in the temperature range 1403~K to 1273~K YMnO$_{3}$ exists in the $P6_{3}/mmc$ phase.

\begin{figure}[b!]
		\includegraphics[width=1.0\columnwidth]{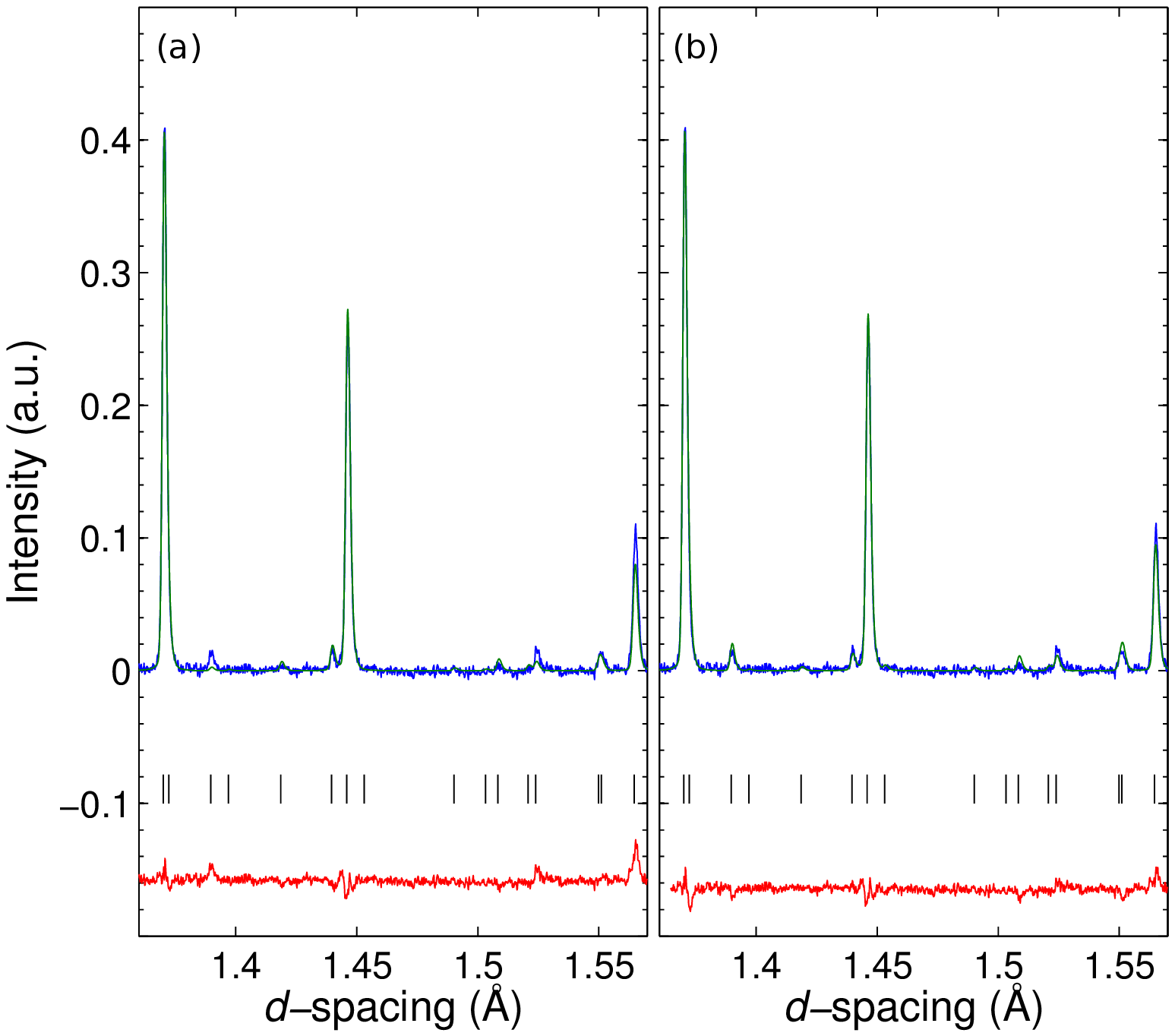}
	\caption{\label{fig:fig5}(Colour Online) Portions of the Rietveld refinement plots for the 1243~K dataset in (a) $P6_{3}/mcm$ and (b) $P6_{3}cm$ models with equal scales. The blue line is the raw data, the green line the Rietveld fit, the red line the difference profile.}
\end{figure}
\begin{figure}[b!]
		\includegraphics[width=1.0\columnwidth]{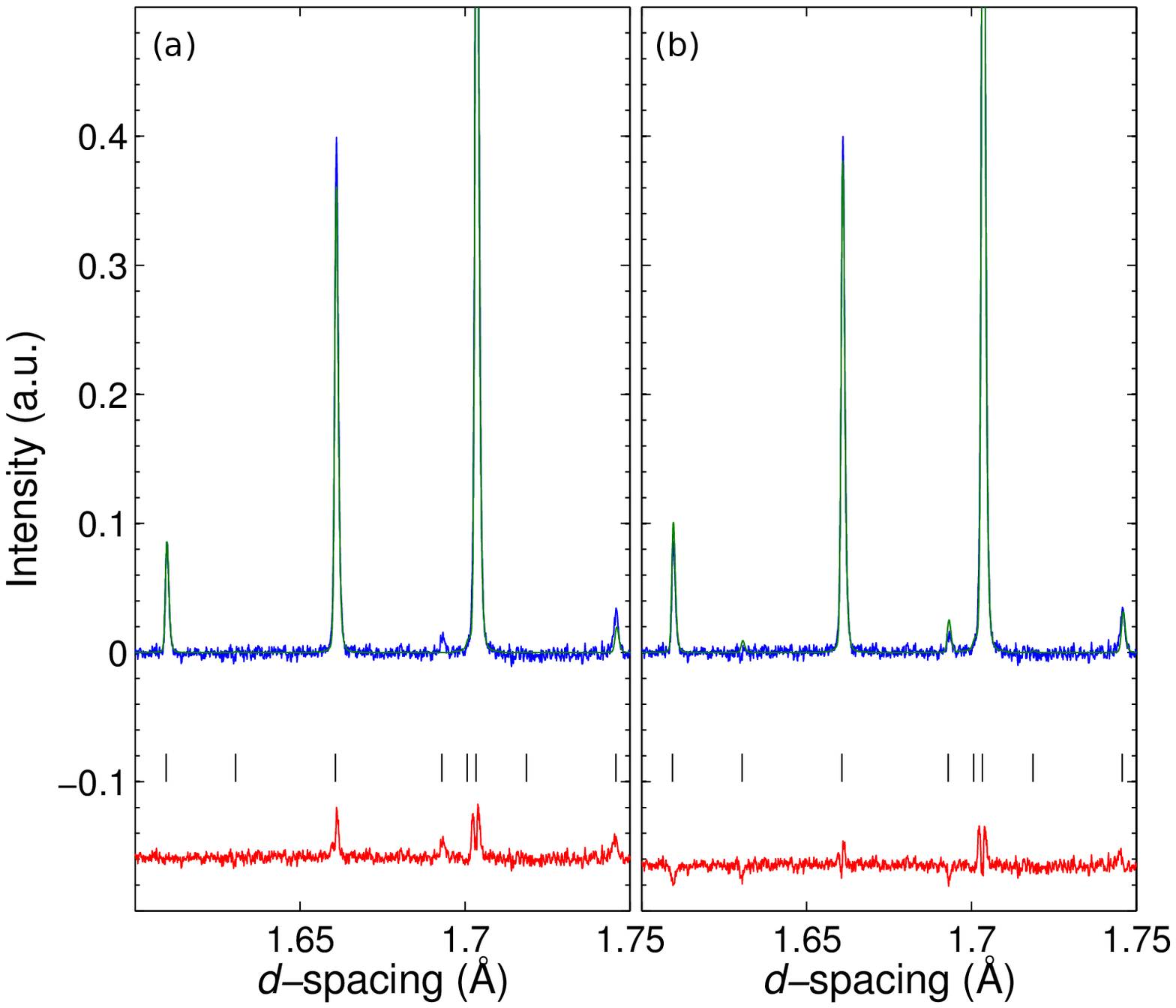}
	\caption{\label{fig:fig6}(Colour Online) Portions of the Rietveld refinement plots for the 1243~K dataset in (a) $P6_{3}/mcm$ and (b) $P6_{3}cm$ models with equal scales. The blue line is the raw data, the green line the Rietveld fit, the red line the difference profile.}
\end{figure}
\subsection{The Unit Cell Tripled Phases}
The dataset immediately below the tripling transition was refined against $P6_{3}cm$ and $P6_{3}/mcm$ models to check for the existence of the previously proposed intermediate phase. Both refinements followed the same strategy with the same time-of-flight range, background type and number of background coefficients. Individual isotropic thermal displacement parameters were refined for all atoms.

Figure \ref{fig:fig4} shows the plot resulting from the Rietveld refinement in space group $P6_{3}cm$ and Figures \ref{fig:fig5} and \ref{fig:fig6} show plots of the comparative refinements over selected \textit{d}-spacing ranges. The $P6_{3}/mcm$ phase proposed by N$\acute{\textrm{e}}$nert et al. \cite{Nenert2007} and Abrahams \cite{Abrahams2009} gave a poorer refinement judged by both \textit{R} factors and by eye. The refined structural parameters and the \textit{wR}$_{p}$ and $\chi^{2}$ values for the $P6_{3}cm$ model are shown in Table \ref{tab:tab3}. The parameters for the $P6_{3}/mcm$ model are included in the supplementary information.

Two specific points may be noted in the $P6_{3}/mcm$ refinement. First, the refined \textit{x} parameter of O1 is within error of $\frac{1}{3}$ (i.e. the allowed displacement does not occur). Second, the thermal displacement parameter, $u_{iso}$, for the equatorial oxygen O3 (corresponding to O4 in the $P6_{3}cm$ model) is almost twice as large as the other two oxygen thermal displacement parameters. Indeed, anisotropic refinement of this atom shows a highly elongated ellipsoid (due to a large $u_{33}$ parameter). This indicates that the continued imposition of the mirror symmetry perpendicular to \textit{c} in the tripled cell is physically unreasonable and the equatorial oxygen plane is in reality tilted. On removal of the mirror plane (i.e. in the polar $P6_{3}cm$ model) both of the equatorial oxide sites and the yttrium sites are allowed to relax their \textit{z}-coordinates leading to significant displacements and more reasonable $u_{iso}$ values. Based on our data, we therefore conclude that the correct space group is $P6_{3}cm$.

\begin{table}[t!]
\caption{\label{tab:tab3}The structural parameters for the 1243~K data refined in $P6_{3}cm$ (No.~185). The lattice parameters are \textit{a}=6.258326(12)~\AA\ and \textit{c}=11.34918(4)~\AA. The refinement gave \textit{wR}$_{p}$=0.0284, $\chi^{2}$=4.854. The variable $u_{iso}$ represents the isotropic thermal displacement parameter for each atomic site.}
\begin{ruledtabular}
\begin{tabular}{c c c c c d}
\multirow{2}{*}{Atom}&Wyckoff&\multirow{2}{*}{\textit{x}}&\multirow{2}{*}{\textit{y}}&\multirow{2}{*}{\textit{z}}&\multicolumn{1}{c}{\multirow{2}{*}{\textit{u}$_{iso}\times$ 100 (\AA$^{2}$)}}\\
&Position&&&&\\\hline
Y1 & 2a&0 & 0 & 0.2639(5) & 2.43(11) \\
Y2 & 4b&$\frac{1}{3}$ & $\frac{2}{3}$ & 0.2399(4) & 2.58(7)\\
Mn & 6c&0.3094(6) & 0 & 0 & 1.62(6) \\
O1 & 6c&0.3297(7) & 0 & 0.1656(4) & 2.55(9) \\
O2 & 6c&0.6621(7) & 0 & 0.3394(4) & 2.98(10) \\
O3 & 2a&0 & 0 & 0.5083(7) & 3.34(17) \\
O4 & 4b&$\frac{1}{3}$ & $\frac{2}{3}$ & 0.0153(5) & 3.18(10) \\
\end{tabular}
\end{ruledtabular}
\end{table}
The Rietveld refinements were performed for all datasets up to the unit cell tripling transition in space group $P6_{3}cm$. The data above the transition was refined in $P6_{3}/mmc$. The refined structural parameters for the room temperature data are shown in Table \ref{tab:tab4} and are in very good agreement with previous single crystal studies \cite{VanAken2004}.

\begin{figure}[t!]
		\includegraphics[width=1.0\columnwidth]{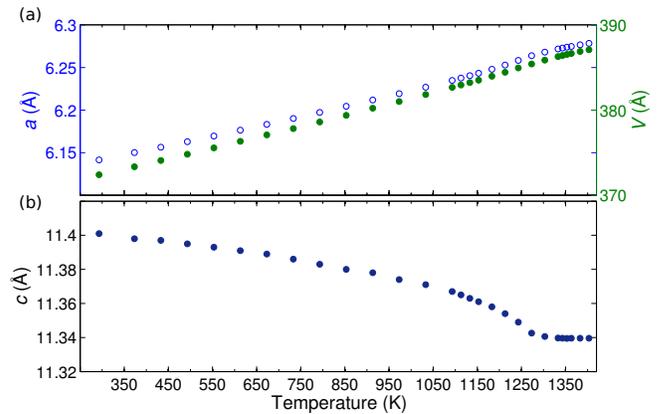}
		\caption{\label{fig:fig7}(Colour Online) (a) The \textit{a}-axis parameter (blue open circles) and unit cell volume (green closed circles) temperature dependences. The \textit{a}-axis parameters from the high temperature phase refinements are scaled by $\sqrt3$ for clarity. (b) The variation of \textit{c}-axis parameter with temperature.}
\end{figure}

\begin{table}[t!]
\caption{\label{tab:tab4}The structural parameters from the Rietveld refinement of the 293 K data in space group $P6_{3}cm$ (No.~185). The unit cell parameters are \textit{a}=6.14151(3) \AA\ and \textit{c}=11.40134(8) \AA. The refinement gave \textit{wR}$_{p}$=0.0330 and $\chi^{2}$=5.885. The variable $u_{iso}$ represents the isotropic thermal displacement parameter for each atomic site.}
\begin{ruledtabular}
\begin{tabular}{l c c c c d}
\multirow{2}{*}{Atom}&Wyckoff&\multirow{2}{*}{\textit{x}}&\multirow{2}{*}{\textit{y}}&\multirow{2}{*}{\textit{z}}&\multicolumn{1}{c}{\multirow{2}{*}{\textit{u}$_{iso}\times$ 100 (\AA$^{2}$)}}\\
&Position&&&&\\\hline
  Y1&2a&0 &  0  &   0.2728(5)    &  1.16(7)\\
  Y2&4b&$\frac{1}{3}$   &   $\frac{2}{3}$ &    0.2325(4)   &   1.30(5)\\     
  Mn1&6c&0.3177(9)&  0   &   0   &   0.80(5)\\    
  O1&6c& 0.3074(4)&  0  &    0.1626(4)   &   1.55(7)\\     
  O2&6c& 0.6427(3)&  0   &   0.3355(4)   &   1.05(6)\\       
  O3&2a& 0 & 0 &   0.4744(6)  &    1.25(10)\\    
  O4&4b&  $\frac{1}{3}$  &  $\frac{2}{3}$  &   0.0169(5)   &   1.42(7) \\
\end{tabular}
\end{ruledtabular}
\end{table}
Figure \ref{fig:fig7} shows the lattice parameters and unit cell volumes extracted from the refinements.
An almost linear temperature dependence of the \textit{a}-axis parameter (the \textit{a}-axis parameter obtained from the refinements of data in the high temperature centrosymmetric phase are scaled by a factor of $\sqrt{3}$) is seen up to about 1100~K where an increase in gradient becomes noticeable. This is followed by a sharp decrease in gradient near 1270~K. The \textit{c}-axis parameter decreases until about 1270~K above which it is roughly constant and the cell volume shows a similar trend to the \textit{a}-axis parameter. The standard deviations of the lattice parameters are of the order of $1.5\times10^{-5}$\AA\ for the \textit{a}-axis and $4\times10^{-5}$\AA\ for the \textit{c}-axis.

The trends in \textit{a} and \textit{V} are better seen by removing a linear term from the temperature dependence.
\begin{figure}[b!]
		\includegraphics[width=1.0\columnwidth]{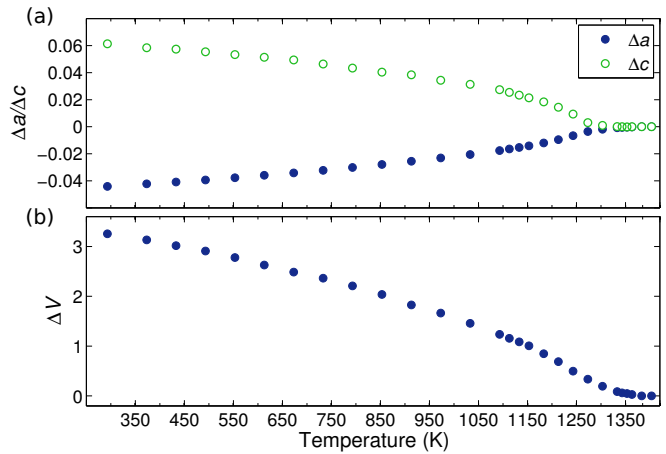}
		\caption{\label{fig:fig8}(Colour Online) (a) The \textit{c}-axis lattice parameter (green open circles) and the \textit{a}-axis parameter (blue closed circles) with a linear temperature dependence (linear fit to highest temperature data) subtracted. (b) The volume of the unit cell with the linear part of the temperature dependence removed.}
\end{figure}
This was done by subtracting a simple linear fit to the highest temperature data (where the dependence is effectively linear) and scaling all data points by the 1403 K value. The results of this are shown in Figure \ref{fig:fig8} and the change in behaviour around 1270 K is clear in all three variables. This is confirmed by inspection of the temperature derivatives of \textit{a} and \textit{c} shown in Figure \ref{fig:fig9}.

\begin{figure}[t!]
		\includegraphics[width=1.0\columnwidth]{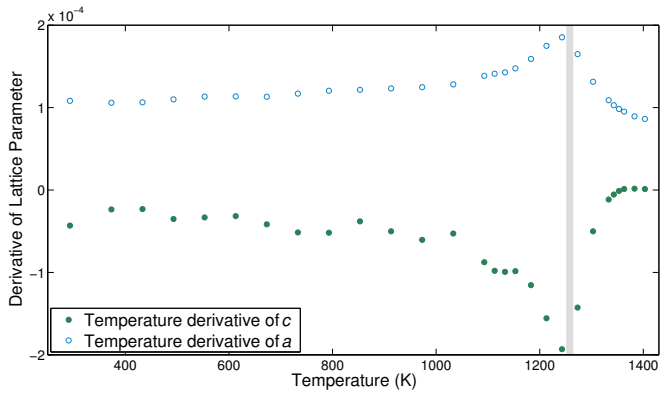}
		\caption{\label{fig:fig9}(Colour Online) The temperature derivatives of the \textit{a} (blue open circles) and \textit{c} (green closed circles) lattice parameters.}

\end{figure}
The tripling of the unit cell directly upon leaving the $P6_{3}/mmc$ phase removes the possibility of a transition through $P6_{3}mc$ as this structure would retain the smaller unit cell. From the lattice parameter derivative data it is clear that the unit cell tripling transition is in the range 1258$\pm$14 K, which confirms the direct evidence from the superlattice peaks presented in Figure \ref{fig:fig3}.

\begin{figure}[t!]
		\includegraphics[width=1.0\columnwidth]{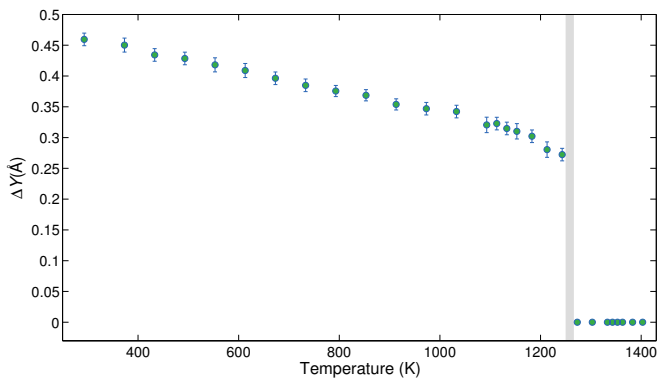}
		\caption{\label{fig:fig10}(Colour Online) The corrugation of the Y layer, $\Delta Y$ is the Y1-Y2 distance in the \textit{c}-direction given by\\ $Y=\Delta(Y1)_{z}+\Delta(Y2)_{z}$.}
\end{figure}

The corrugation of the Y$^{3+}$ layers may also be examined to investigate displacements occurring as the system moves towards the high temperature centrosymmetric phase. This phase has a single Y$^{3+}$ site with $\overline{3}m$ symmetry at fractional coordinates (0,0,$\frac{1}{4}$). When the symmetry is lowered, the mirror plane is lost and two of the six Y$^{3+}$ ions displace `upwards' along \textit{z} (Y1, Wyckoff site 2a) while four displace `downwards' (Y2, Wyckoff site 4b) from the centrosymmetric position. The sum of the \textit{c}-axis displacements, in \AA ngstr\"{o}ms for a single ion from each site, $\Delta Y=\Delta(Y1)_{z}+\Delta(Y2)_{z}$, is shown in Figure \ref{fig:fig10}.
The corrugation increases from zero in the $P6_{3}/mmc$ phase to about 0.27~\AA\ sharply and then increases smoothly with decreasing temperature.
\subsection{Secondary Transition}
\begin{figure}[b!]
		\includegraphics[width=1.0\columnwidth]{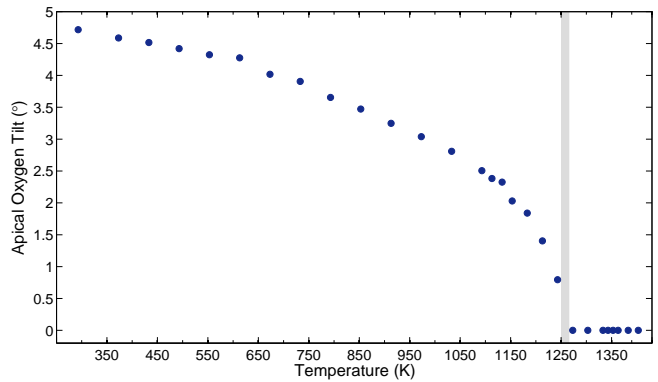}
	\caption{\label{fig:fig11}(Colour Online) The tilt angle calculated for the Mn apical oxygens (O1 and O2) relative to the \textit{c}-axis.}
\end{figure}

In the higher symmetry $P6_{3}/mmc$ phase, the MnO$_{5}$ trigonal bipyramids are untilted. It can be seen from Figure \ref{fig:fig11} that the evolution of apical tilt angle (calculated by taking the angle of O1-O2 to the \textit{c}-axis direction) with decreasing temperature is relatively smooth, being fixed to zero by symmetry at and above 1273 K. 
\begin{figure}[b!]
		\includegraphics[width=1.0\columnwidth]{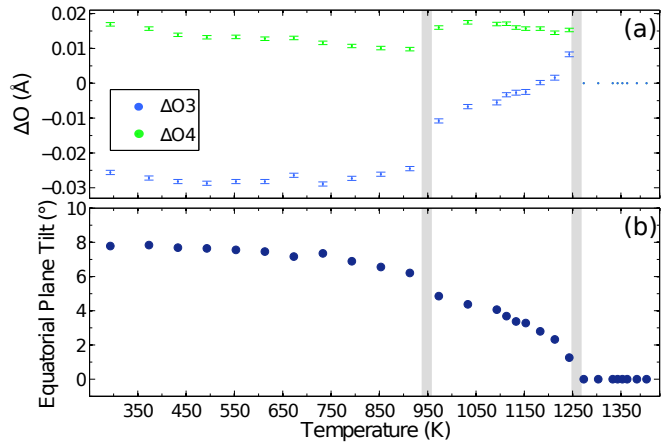}
	\caption{\label{fig:fig12}(Colour Online) (a)The displacements of O3 (blue) and O4 (green) from their centrosymmetric positions in the \textit{z} direction. (b) The tilt of the equatorial oxygen plane with respect to the \textit{ab}-plane.}
\end{figure}
Inspection of the O3 and O4 \textit{z} coordinates, shown in Figure \ref{fig:fig12}(a), shows a feature around 913 K with both \textit{z}-coordinates suddenly decreasing. The displacement of O3, in particular, increases markedly with decreasing temperature, moving below the Mn ion for $T\!\leq\,$853 K. Figure \ref{fig:fig12}(b) shows the tilt of the equatorial oxygen plane with temperature (angle of O3-O4 to the \textit{ab}-plane), there is a sharp increase at $T\!\leq\,$913~K. The sudden decrease in O3(\textit{z}) and O4(\textit{z}) and increase in the equatorial oxygen plane tilt would correspond to a sharp decrease in polarization. This would account for the peak in pyroelectric current seen by Ismailzade and Kizhaev \cite{Ismailzade1965} as pyroelectric current is proportional to $\frac{\partial{P}}{\partial{T}}$ for constant heating rate. The resistivity data of Katsufuji \textit{et al.}\cite{Katsufuji2001} and Choi \textit{et al.}\cite{Choi2010} also show a crossover in resistivity behaviour in this temperature region. 

\begin{figure}[t!]
		\includegraphics[width=1.0\columnwidth]{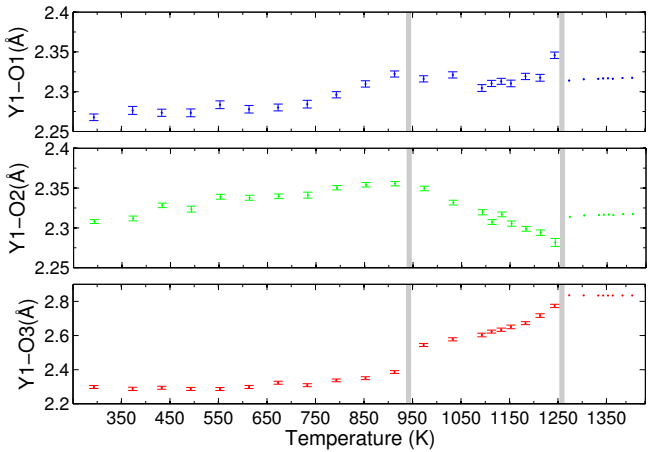}
		\caption{\label{fig:fig13}(Colour Online) The Y1-O bond lengths determined from the Rietveld refinements.}
\end{figure}
\begin{figure}[t!]
		\includegraphics[width=0.8\columnwidth]{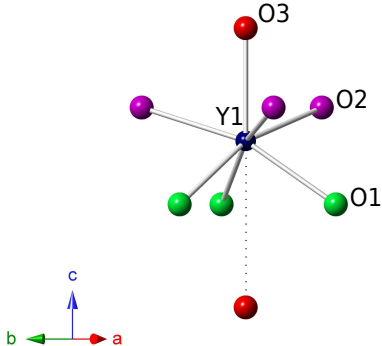}
	\caption{\label{fig:fig14}(Colour Online) The coordination of the Y1 ion in the low-temperature $P6_{3}cm$ structure. The bond lengths at room temperature obtained from the Rietveld refinement are Y1-O1=2.267(4)~\AA, Y1-O2=2.308(2)~\AA\ and Y1-O3=2.298(8)~\AA. The distance to the other O3 ion (indicated by the dotted line) is 3.402(8)~\AA.}
\end{figure}
The Y-O bond lengths shown in Figure~\ref{fig:fig13} vary smoothly within error bars except for around 913~K where there is a sudden decrease in the Y1-O3 bond length. This corresponds to the bonding change seen by Kim \textit{et al} \cite{Kim2009}. These authors observed hybridisation of Y1 and O3 between 910~K and 930~K by Maximum Entropy Method (MEM) analysis of the electron density from synchrotron x-ray data. The O3 coordinates the Y1 site apically whereas the O1 and O2 coordinate it in the equatorial direction as shown in Figure \ref{fig:fig14}.
The concordance of the structural changes seen in the present study with the pyroelectric current, resistivity and x-ray diffraction data strongly suggests that there is an isosymmetric phase transition taking place at $\approx\,$920 K.

\begin{figure}[t!]
		\includegraphics[width=1.0\columnwidth]{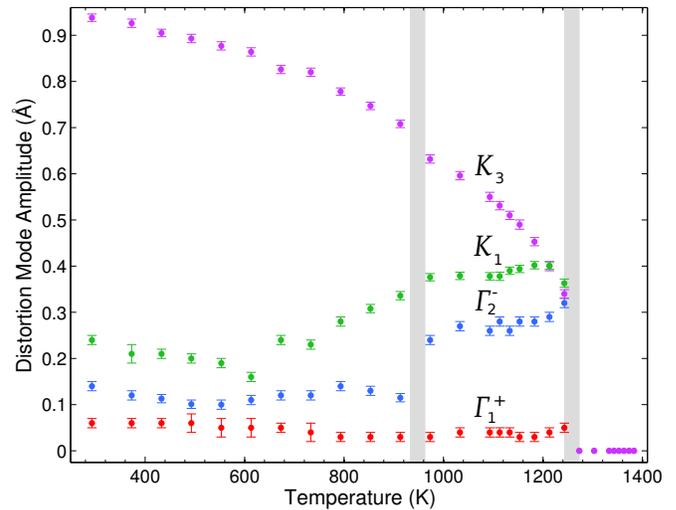}
\caption{\label{fig:fig15}(Colour Online) The amplitudes of the distortion modes calculated using Amplimodes \cite{Orobengoa2009,Perez-Mato2010}. Above 1243~K all mode amplitudes are zero and are included for completeness only.}
	
\end{figure}

The program Amplimodes \cite{Orobengoa2009,Perez-Mato2010} was used to examine the magnitude of the $\mathit{\Gamma}_{1}^{+}$, $\mathit{\Gamma}_{2}^{-}$, $K_{1}$ and $K_{3}$ distortion modes of $P6_{3}/mmc$. The results are shown in Figure \ref{fig:fig15}. The nature of these distortion modes can also be examined and visualised using Isodisplace\cite{Campbell2006}. $\mathit{\Gamma}_{1}^{+}$ corresponds to a symmetric breathing mode with the order parameter being the change in \textit{z} coordinate of the apical oxygen O1. $\mathit{\Gamma}_{2}^{-}$  involves a polar displacement of the ions along the \textit{c}-axis leading ultimately to the space group $P6_{3}mc$. $K_{1}$ leads to the space group $P6_{3}/mcm$ by allowing only the O-Mn-O axis to displace in the \textit{ab}-plane, tripling the unit cell. $K_{3}$ is the antiferrodistortive mode leading to $P6_{3}cm$ by a tilt of the MnO$_{5}$ bipyramids and antiparallel displacements of the Y$^{3+}$ cations leading to unit cell tripling.

\begin{figure}[t!]
		\includegraphics[width=1.0\columnwidth]{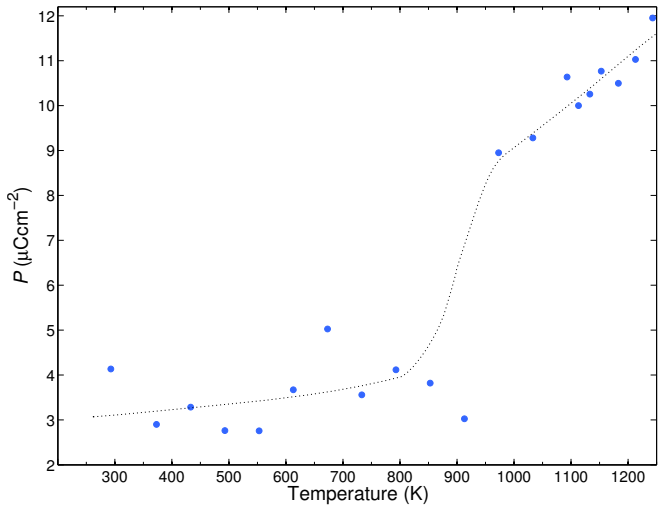}
	\caption{\label{fig:fig16}(Colour Online) The polarization of YMnO$_{3}$ estimated using a simple ionic model with the nuclear positions obtained from the Rietveld refinements. The dotted line is a guide to the eye.}
\end{figure}

It is clear that the $K_{3}$ mode is dominant in general although the $K_{1}$ mode amplitude is slightly larger for the 1243 K dataset (we note that this results almost exclusively from a significant displacement of the Mn \textit{x}-coordinate (see supplementary material): since the Mn atom has the smallest neutron scattering length, this is the least well-determined positional parameter in this study). This dataset was refined in both $P6_{3}/mcm$ and $P6_{3}cm$ space groups as already stated and $P6_{3}cm$ gave the better fit. At the secondary transition ($\approx$920 K), the amplitude of the $\mathit{\Gamma}_{2}^{-}$ mode decreases abruptly and the $K_{3}$ mode shows a small step increase. This would (as a partial contribution to the polarization of the $P6_{3}cm$ structure) cause a sharp decrease in the polarization and corresponds to the behaviour seen in O3 and O4 parameters. The polarization, $P$, was estimated for the refined structures using a simple ionic model, \[P=\sum_{i}\frac{\Delta c_{i} Q_{i}em_{i}}{V},\] where $\Delta c_{i}$ is the displacement of the site from the centrosymmetric position in \AA ngstr\"{o}ms, $Q_{i}$ is the ionic charge, $e$ the electron charge, the site multiplicity is denoted by $m_{i}$ and the unit cell volume by $V$. The results of this estimate are shown in Figure \ref{fig:fig16}.

\section{Conclusions}
In summary, our data confirm that there is no intermediate phase (of the proposed $P6_{3}/mcm$ symmetry or any other candidate symmetries) between the high temperature, paraelectric $P6_{3}/mmc$ phase and the unit-cell tripled polar $P6_{3}cm$ phase. This transition occurs between 1243 and 1273 K and is driven primarily by a non-polar displacement mode of $K_{3}$ symmetry in agreement with earlier theoretical works \cite{Fennie2005}. Hexagonal YMnO$_{3}$ is therefore an improper ferroelectric, with the antiferrodistortive $K_{3}$ mode triggering a weaker, polar distortion of $\mathit{\Gamma}_{2}^{-}$ type. Although there is no clearly defined `intermediate' phase of differing crystallographic symmetry, our data provide subtle, but compelling evidence of a secondary isosymmetric transition in the $P6_{3}cm$ regime, at around 920 K. This transition involves polar displacements of the Mn-O equatorial planes, and may be related to an electronic transition involving hybridisation of the Y1-O3 bond. Perhaps surprisingly, this leads to a \textit{decreased} polarization but nevertheless does agree with the various anomalies in physical properties reported around this temperature.

\begin{acknowledgments}
 We thank the EPSRC and STFC for funding, K.E.~Johnston for experimental assistance and F.D.~Morrison for discussions.
\end{acknowledgments}

\clearpage

{\section*{APPENDIX: Supplementary Information}
%\vspace*{10 mm}
\begin{figure}[!h]
		\includegraphics[width=1.0\columnwidth]{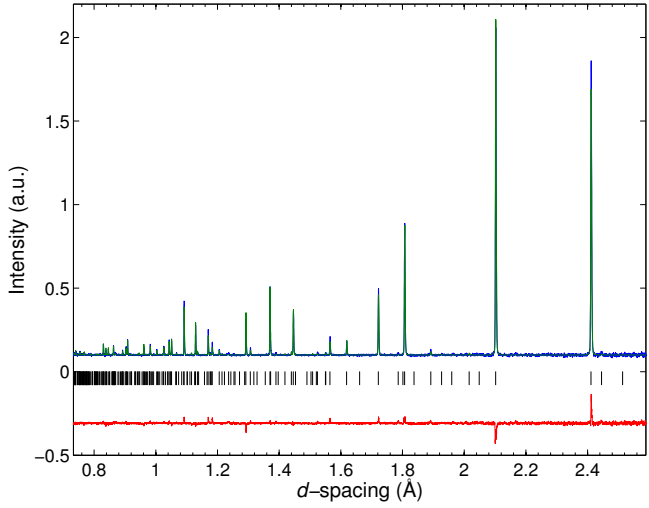}
	\caption{\label{fig:fig17}The results of the refinement of the 1243~K dataset with the $P6_{3}/mcm$ model. The raw data is shown in blue, the Rietveld fit in green and the difference profile in red.}
	\end{figure}
\begin{table}[!h]
\vspace*{15mm}
\caption{\label{tab:tab5}The structural parameters from the Rietveld refinement of the 293~K data in space group $P6_{3}cm$ (No.~185). The unit cell parameters are \textit{a}=6.258342(15) \AA\ and \textit{c}=11.34917(5) \AA. The refinement gave \textit{wR}$_{p}$=0.0335 and $\chi^{2}$=6.750.}
\begin{ruledtabular}
\begin{tabular}{l c c c c d}
\multirow{2}{*}{Atom}&Wyckoff&\multirow{2}{*}{\textit{x}}&\multirow{2}{*}{\textit{y}}&\multirow{2}{*}{\textit{z}}&\multicolumn{1}{c}{\multirow{2}{*}{\textit{u}$_{iso}\times$ 100 (\AA$^{2}$)}}\\
&Position&&&&\\\hline
  Y1&2b&0&0&$\frac{1}{2}$&3.78(15)\\
 Y2&4d&$\frac{1}{3}$&$\frac{2}{3}$&$\frac{1}{2}$&2.84(7)\\     
  Mn&6g&0.3067(6)&0&$\frac{1}{4}$&1.82(7)  \\
  O1&12k&0.3344(4)&0&0.08845(9)&2.62(4)\\
 O2&2a&0&0&$\frac{1}{4}$&2.40(14)\\
O3&4c&$\frac{1}{3}$&$\frac{2}{3}$&$\frac{1}{4}$&4.79(12)\\
\end{tabular}
\end{ruledtabular}
\end{table}
\newpage
%\
\begin{figure}[!h]
\vspace*{5 mm}
		\includegraphics[width=1.0\columnwidth]{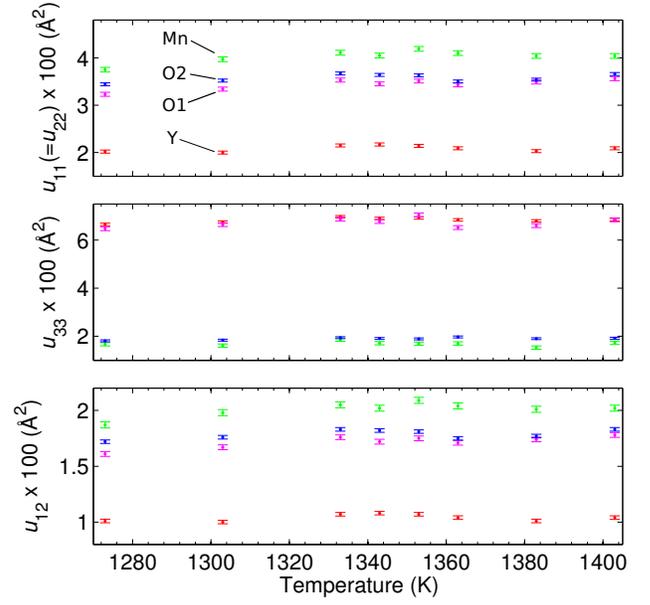}
	\caption{\label{fig:fig18}The anisotropic thermal displacement parameters obtained from the Rietveld refinements of datasets at and above 1273~K with the $P6_{3}/mmc$ model.}
\end{figure}
\begin{figure}[!h]
%\vspace*{14 mm}
		\includegraphics[width=1.0\columnwidth]{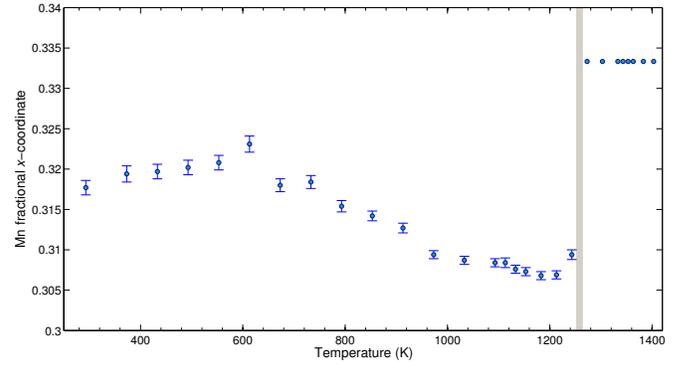}
	\caption{\label{fig:fig19}The Mn \textit{x}-coordinate as a function of temperature. The position is fixed to $\frac{1}{3}$ in the $P6_{3}/mmc$ phase and is shown for completeness.}
\end{figure}
\newpage
\begin{table*}[!t]
\vspace*{20 mm}
\caption{\label{tab:tab6}The results of the Rietveld refinement of the 1273~K dataset in the space group $P6_{3}mc$ (No.~186). The lattice parameters are \textit{a} = 3.616572(6) and \textit{c} = 11.34308(3). The refinement gave \textit{wR}$_{p}$=0.0233, $\chi^{2}$=3.138.}
\begin{ruledtabular}
\begin{tabular}{c c c c c d d d}
\multirow{2}{*}{Atom}&Wyckoff&\multirow{2}{*}{\textit{x}}&\multirow{2}{*}{\textit{y}}&\multirow{2}{*}{\textit{z}}&\multicolumn{1}{c}{\multirow{2}{*}{$u_{11}/u_{22}\times$ 100 (\AA$^{2}$)}}&\multicolumn{1}{c}{\multirow{2}{*}{$u_{33}\times$ 100 (\AA$^{2}$)}}&\multicolumn{1}{c}{\multirow{2}{*}{$u_{12}\times$ 100 (\AA$^{2}$)}}\\
&Position&&&&&&\\\hline
Y & 2a&0&0&0&2.03(2)&6.52(7)&1.016(12)\\
Mn & 2b&$\frac{1}{3}$&$\frac{2}{3}$&0.2522(13)&3.68(5)&1.56(8)&1.84(2)\\
O1 & 2a&0&0&0.2535(11)&3.23(4)&6.55(10)&1.616(17)\\
O2 & 2b&$\frac{1}{3}$&$\frac{2}{3}$&0.0895(9)&3.4(3)&2.0(3)&1.76(17)\\
O3 & 2b&$\frac{1}{3}$&$\frac{2}{3}$&0.9185(9)&3.4(3)&1.7(3)&1.69(17)\\
\end{tabular}
\end{ruledtabular}
\end{table*}
\begin{table*}[!t]
\vspace*{-110 mm}
\caption{\label{tab:tab7}The structural parameters for the 1273~K data refined in $P6_{3}/mmc$ (No.~194). The lattice parameters are \textit{a} = 3.616572(6) and \textit{c} = 11.34308(3). The refinement gave \textit{wR}$_{p}$=0.0233, $\chi^{2}$=3.138.}
\begin{ruledtabular}
\begin{tabular}{c c c c c d d d}
\multirow{2}{*}{Atom}&Wyckoff&\multirow{2}{*}{\textit{x}}&\multirow{2}{*}{\textit{y}}&\multirow{2}{*}{\textit{z}}&\multicolumn{1}{c}{\multirow{2}{*}{$u_{11}/u_{22}\times$ 100 (\AA$^{2}$)}}&\multicolumn{1}{c}{\multirow{2}{*}{$u_{33}\times$ 100 (\AA$^{2}$)}}&\multicolumn{1}{c}{\multirow{2}{*}{$u_{12}\times$ 100 (\AA$^{2}$)}}\\
&Position&&&&&&\\\hline
Y1 & 2a&0 &0&0&2.03(3)&6.65(5)&1.015(16)\\
Mn & 2c&$\frac{1}{3}$&$\frac{2}{3}$&$\frac{1}{4}$&3.75(5)&1.67(7)&1.87(2)\\
O1 & 2b&0& 0&$\frac{1}{4}$&3.24(4)&6.47(7)&1.618(2)\\
O2 & 4f&$\frac{1}{3}$&$\frac{2}{3}$&0.08557(7)&3.44(3)&1.81(4)&1.722(15)\\
\end{tabular}
\end{ruledtabular}
\end{table*}

\begin{thebibliography}{24}%
\makeatletter
\providecommand \@ifxundefined [1]{%
 \@ifx{#1\undefined}
}%
\providecommand \@ifnum [1]{%
 \ifnum #1\expandafter \@firstoftwo
 \else \expandafter \@secondoftwo
 \fi
}%
\providecommand \@ifx [1]{%
 \ifx #1\expandafter \@firstoftwo
 \else \expandafter \@secondoftwo
 \fi
}%
\providecommand \natexlab [1]{#1}%
\providecommand \enquote  [1]{``#1''}%
\providecommand \bibnamefont  [1]{#1}%
\providecommand \bibfnamefont [1]{#1}%
\providecommand \citenamefont [1]{#1}%
\providecommand \href@noop [0]{\@secondoftwo}%
\providecommand \href [0]{\begingroup \@sanitize@url \@href}%
\providecommand \@href[1]{\@@startlink{#1}\@@href}%
\providecommand \@@href[1]{\endgroup#1\@@endlink}%
\providecommand \@sanitize@url [0]{\catcode `\\12\catcode `\$12\catcode
  `\&12\catcode `\#12\catcode `\^12\catcode `\_12\catcode `\%12\relax}%
\providecommand \@@startlink[1]{}%
\providecommand \@@endlink[0]{}%
\providecommand \url  [0]{\begingroup\@sanitize@url \@url }%
\providecommand \@url [1]{\endgroup\@href {#1}{\urlprefix }}%
\providecommand \urlprefix  [0]{URL }%
\providecommand \Eprint [0]{\href }%
\@ifxundefined \urlstyle {%
  \providecommand \doi  [0]{\begingroup \@sanitize@url \@doi}%
  \providecommand \@doi [1]{\endgroup \@@startlink {\doibase
  #1}doi:\discretionary {}{}{}#1\@@endlink }%
}{%
  \providecommand \doi  [0]{doi:\discretionary{}{}{}\begingroup
  \urlstyle{rm}\Url }%
}%
\providecommand \doibase [0]{http://dx.doi.org/}%
\providecommand \Doi [0]{\begingroup \@sanitize@url \@Doi }%
\providecommand \@Doi  [1]{\endgroup\@@startlink{\doibase#1}\@@Doi}%
\providecommand \@@Doi [1]{#1\@@endlink}%
\providecommand \selectlanguage [0]{\@gobble}%
\providecommand \bibinfo  [0]{\@secondoftwo}%
\providecommand \bibfield  [0]{\@secondoftwo}%
\providecommand \translation [1]{[#1]}%
\providecommand \BibitemOpen [0]{}%
\providecommand \bibitemStop [0]{}%
\providecommand \bibitemNoStop [0]{.\EOS\space}%
\providecommand \EOS [0]{\spacefactor3000\relax}%
\providecommand \BibitemShut  [1]{\csname bibitem#1\endcsname}%
%</preamble>
\bibitem [{\citenamefont {Lee}\ \emph {et~al.}(2008)\citenamefont {Lee},
  \citenamefont {Pirogov}, \citenamefont {Kang}, \citenamefont {Jang},
  \citenamefont {Yonemura}, \citenamefont {Kamiyama}, \citenamefont {Cheong},
  \citenamefont {Gozzo}, \citenamefont {Shin}, \citenamefont {Kimura},
  \citenamefont {Noda},\ and\ \citenamefont {Park}}]{Lee2008}%
  \BibitemOpen
  \bibfield  {author} {\bibinfo {author} {\bibfnamefont {S.}~\bibnamefont
  {Lee}}, \bibinfo {author} {\bibfnamefont {A.}~\bibnamefont {Pirogov}},
  \bibinfo {author} {\bibfnamefont {M.~S.}\ \bibnamefont {Kang}}, \bibinfo
  {author} {\bibfnamefont {K.~H.}\ \bibnamefont {Jang}}, \bibinfo {author}
  {\bibfnamefont {M.}~\bibnamefont {Yonemura}}, \bibinfo {author}
  {\bibfnamefont {T.}~\bibnamefont {Kamiyama}}, \bibinfo {author}
  {\bibfnamefont {S.~W.}\ \bibnamefont {Cheong}}, \bibinfo {author}
  {\bibfnamefont {F.}~\bibnamefont {Gozzo}}, \bibinfo {author} {\bibfnamefont
  {N.}~\bibnamefont {Shin}}, \bibinfo {author} {\bibfnamefont {H.}~\bibnamefont
  {Kimura}}, \bibinfo {author} {\bibfnamefont {Y.}~\bibnamefont {Noda}}, \ and\
  \bibinfo {author} {\bibfnamefont {J.~G.}\ \bibnamefont {Park}},\ }\href@noop
  {} {\bibfield  {journal} {\bibinfo  {journal} {Nature},\ }\textbf {\bibinfo
  {volume} {451}},\ \bibinfo {pages} {805} (\bibinfo {year}
  {2008})}\BibitemShut {NoStop}%
\bibitem [{\citenamefont {Choi}\ \emph {et~al.}(2010)\citenamefont {Choi},
  \citenamefont {Horibe}, \citenamefont {Yi}, \citenamefont {Choi},
  \citenamefont {Wu},\ and\ \citenamefont {Cheong}}]{Choi2010}%
  \BibitemOpen
  \bibfield  {author} {\bibinfo {author} {\bibfnamefont {T.}~\bibnamefont
  {Choi}}, \bibinfo {author} {\bibfnamefont {Y.}~\bibnamefont {Horibe}},
  \bibinfo {author} {\bibfnamefont {H.~T.}\ \bibnamefont {Yi}}, \bibinfo
  {author} {\bibfnamefont {Y.~J.}\ \bibnamefont {Choi}}, \bibinfo {author}
  {\bibfnamefont {W.~D.}\ \bibnamefont {Wu}}, \ and\ \bibinfo {author}
  {\bibfnamefont {S.~W.}\ \bibnamefont {Cheong}},\ }\href@noop {} {\bibfield
  {journal} {\bibinfo  {journal} {Nature Materials},\ }\textbf {\bibinfo
  {volume} {9}},\ \bibinfo {pages} {253} (\bibinfo {year} {2010})}\BibitemShut
  {NoStop}%
\bibitem [{\citenamefont {Zhou}\ \emph {et~al.}(2006)\citenamefont {Zhou},
  \citenamefont {Goodenough}, \citenamefont {Gallardo-Amores}, \citenamefont
  {{Mor\'an}}, \citenamefont {Alario-Franco},\ and\ \citenamefont
  {Caudillo}}]{Zhou2006}%
  \BibitemOpen
  \bibfield  {author} {\bibinfo {author} {\bibfnamefont {J.-S.}\ \bibnamefont
  {Zhou}}, \bibinfo {author} {\bibfnamefont {J.~B.}\ \bibnamefont
  {Goodenough}}, \bibinfo {author} {\bibfnamefont {J.~M.}\ \bibnamefont
  {Gallardo-Amores}}, \bibinfo {author} {\bibfnamefont {E.}~\bibnamefont
  {{Mor\'an}}}, \bibinfo {author} {\bibfnamefont {M.~A.}\ \bibnamefont
  {Alario-Franco}}, \ and\ \bibinfo {author} {\bibfnamefont {R.}~\bibnamefont
  {Caudillo}},\ }\href@noop {} {\bibfield  {journal} {\bibinfo  {journal}
  {Phys. Rev. B},\ }\textbf {\bibinfo {volume} {74}},\ \bibinfo {pages}
  {014422} (\bibinfo {year} {2006})}\BibitemShut {NoStop}%
\bibitem [{\citenamefont {Abrahams}(2001)}]{Abrahams2001}%
  \BibitemOpen
  \bibfield  {author} {\bibinfo {author} {\bibfnamefont {S.~C.}\ \bibnamefont
  {Abrahams}},\ }\href@noop {} {\bibfield  {journal} {\bibinfo  {journal} {Acta
  Crystallographica Section B-structural Science},\ }\textbf {\bibinfo {volume}
  {57}},\ \bibinfo {pages} {485} (\bibinfo {year} {2001})}\BibitemShut
  {NoStop}%
\bibitem [{\citenamefont {Chatterji}\ \emph {et~al.}(2007)\citenamefont
  {Chatterji}, \citenamefont {Ghosh}, \citenamefont {Singh}, \citenamefont
  {Regnault},\ and\ \citenamefont {Rheinst\"adter}}]{Chatterji2007}%
  \BibitemOpen
  \bibfield  {author} {\bibinfo {author} {\bibfnamefont {T.}~\bibnamefont
  {Chatterji}}, \bibinfo {author} {\bibfnamefont {S.}~\bibnamefont {Ghosh}},
  \bibinfo {author} {\bibfnamefont {A.}~\bibnamefont {Singh}}, \bibinfo
  {author} {\bibfnamefont {L.~P.}\ \bibnamefont {Regnault}}, \ and\ \bibinfo
  {author} {\bibfnamefont {M.}~\bibnamefont {Rheinst\"adter}},\ }\href@noop {}
  {\bibfield  {journal} {\bibinfo  {journal} {Phys. Rev. B},\ }\textbf
  {\bibinfo {volume} {76}},\ \bibinfo {pages} {144406} (\bibinfo {year}
  {2007})}\BibitemShut {NoStop}%
\bibitem [{\citenamefont {Fennie}\ and\ \citenamefont
  {Rabe}(2005)}]{Fennie2005}%
  \BibitemOpen
  \bibfield  {author} {\bibinfo {author} {\bibfnamefont {C.~J.}\ \bibnamefont
  {Fennie}}\ and\ \bibinfo {author} {\bibfnamefont {K.~M.}\ \bibnamefont
  {Rabe}},\ }\href@noop {} {\bibfield  {journal} {\bibinfo  {journal} {Phys.
  Rev. B},\ }\textbf {\bibinfo {volume} {72}},\ \bibinfo {pages} {100103}
  (\bibinfo {year} {2005})}\BibitemShut {NoStop}%
\bibitem [{\citenamefont {Van~Aken}\ \emph {et~al.}(2004)\citenamefont
  {Van~Aken}, \citenamefont {Palstra}, \citenamefont {Filippetti},\ and\
  \citenamefont {Spaldin}}]{VanAken2004}%
  \BibitemOpen
  \bibfield  {author} {\bibinfo {author} {\bibfnamefont {B.~B.}\ \bibnamefont
  {Van~Aken}}, \bibinfo {author} {\bibfnamefont {T.~T.~M.}\ \bibnamefont
  {Palstra}}, \bibinfo {author} {\bibfnamefont {A.}~\bibnamefont {Filippetti}},
  \ and\ \bibinfo {author} {\bibfnamefont {N.~A.}\ \bibnamefont {Spaldin}},\
  }\href@noop {} {\bibfield  {journal} {\bibinfo  {journal} {Nature
  Materials},\ }\textbf {\bibinfo {volume} {3}},\ \bibinfo {pages} {164}
  (\bibinfo {year} {2004})}\BibitemShut {NoStop}%
\bibitem [{\citenamefont {Khomskii}(2006)}]{Khomskii2006}%
  \BibitemOpen
  \bibfield  {author} {\bibinfo {author} {\bibfnamefont {D.~I.}\ \bibnamefont
  {Khomskii}},\ }\href@noop {} {\bibfield  {journal} {\bibinfo  {journal}
  {Journal of Magnetism and Magnetic Materials},\ }\textbf {\bibinfo {volume}
  {306}},\ \bibinfo {pages} {1} (\bibinfo {year} {2006})}\BibitemShut {NoStop}%
\bibitem [{\citenamefont {Ismailzade}\ and\ \citenamefont
  {Kizhaev}(1965)}]{Ismailzade1965}%
  \BibitemOpen
  \bibfield  {author} {\bibinfo {author} {\bibfnamefont {I.~G.}\ \bibnamefont
  {Ismailzade}}\ and\ \bibinfo {author} {\bibfnamefont {S.~A.}\ \bibnamefont
  {Kizhaev}},\ }\href@noop {} {\bibfield  {journal} {\bibinfo  {journal}
  {Soviet Physics Solid State},\ }\textbf {\bibinfo {volume} {7}},\ \bibinfo
  {pages} {236} (\bibinfo {year} {1965})}\BibitemShut {NoStop}%
\bibitem [{\citenamefont {{\L ukaszewicz}}\ and\ \citenamefont
  {{Karut-Kalici$\acute{\textrm{n}}$ska}}(1974)}]{Luka1974}%
  \BibitemOpen
  \bibfield  {author} {\bibinfo {author} {\bibfnamefont {K.}~\bibnamefont {{\L
  ukaszewicz}}}\ and\ \bibinfo {author} {\bibfnamefont {J.}~\bibnamefont
  {{Karut-Kalici$\acute{\textrm{n}}$ska}}},\ }\href@noop {} {\bibfield
  {journal} {\bibinfo  {journal} {Ferroelectrics},\ }\textbf {\bibinfo {volume}
  {7}},\ \bibinfo {pages} {81} (\bibinfo {year} {1974})}\BibitemShut {NoStop}%
\bibitem [{\citenamefont {Katsufuji}\ \emph {et~al.}(2001)\citenamefont
  {Katsufuji}, \citenamefont {Mori}, \citenamefont {Masaki}, \citenamefont
  {Moritomo}, \citenamefont {Yamamoto},\ and\ \citenamefont
  {Takagi}}]{Katsufuji2001}%
  \BibitemOpen
  \bibfield  {author} {\bibinfo {author} {\bibfnamefont {T.}~\bibnamefont
  {Katsufuji}}, \bibinfo {author} {\bibfnamefont {S.}~\bibnamefont {Mori}},
  \bibinfo {author} {\bibfnamefont {M.}~\bibnamefont {Masaki}}, \bibinfo
  {author} {\bibfnamefont {Y.}~\bibnamefont {Moritomo}}, \bibinfo {author}
  {\bibfnamefont {N.}~\bibnamefont {Yamamoto}}, \ and\ \bibinfo {author}
  {\bibfnamefont {H.}~\bibnamefont {Takagi}},\ }\href@noop {} {\bibfield
  {journal} {\bibinfo  {journal} {Phys. Rev. B},\ }\textbf {\bibinfo {volume}
  {64}},\ \bibinfo {pages} {104419} (\bibinfo {year} {2001})}\BibitemShut
  {NoStop}%
\bibitem [{\citenamefont {Katsufuji}\ \emph {et~al.}(2002)\citenamefont
  {Katsufuji}, \citenamefont {Masaki}, \citenamefont {Machida}, \citenamefont
  {Moritomo}, \citenamefont {Kato}, \citenamefont {Nishibori}, \citenamefont
  {Takata}, \citenamefont {Sakata}, \citenamefont {Ohoyama}, \citenamefont
  {Kitazawa},\ and\ \citenamefont {Takagi}}]{Katsufuji2002}%
  \BibitemOpen
  \bibfield  {author} {\bibinfo {author} {\bibfnamefont {T.}~\bibnamefont
  {Katsufuji}}, \bibinfo {author} {\bibfnamefont {M.}~\bibnamefont {Masaki}},
  \bibinfo {author} {\bibfnamefont {A.}~\bibnamefont {Machida}}, \bibinfo
  {author} {\bibfnamefont {M.}~\bibnamefont {Moritomo}}, \bibinfo {author}
  {\bibfnamefont {K.}~\bibnamefont {Kato}}, \bibinfo {author} {\bibfnamefont
  {E.}~\bibnamefont {Nishibori}}, \bibinfo {author} {\bibfnamefont
  {M.}~\bibnamefont {Takata}}, \bibinfo {author} {\bibfnamefont
  {M.}~\bibnamefont {Sakata}}, \bibinfo {author} {\bibfnamefont
  {K.}~\bibnamefont {Ohoyama}}, \bibinfo {author} {\bibfnamefont
  {K.}~\bibnamefont {Kitazawa}}, \ and\ \bibinfo {author} {\bibfnamefont
  {H.}~\bibnamefont {Takagi}},\ }\href@noop {} {\bibfield  {journal} {\bibinfo
  {journal} {Phys. Rev. B},\ }\textbf {\bibinfo {volume} {66}},\ \bibinfo
  {pages} {134434} (\bibinfo {year} {2002})}\BibitemShut {NoStop}%
\bibitem [{\citenamefont {N$\acute{\textrm{e}}$nert}\ \emph
  {et~al.}(2007){\natexlab{a}}\citenamefont {N$\acute{\textrm{e}}$nert},
  \citenamefont {Ren}, \citenamefont {Stokes},\ and\ \citenamefont
  {Palstra}}]{Nenert2005}%
  \BibitemOpen
  \bibfield  {author} {\bibinfo {author} {\bibfnamefont {G.}~\bibnamefont
  {N$\acute{\textrm{e}}$nert}}, \bibinfo {author} {\bibfnamefont
  {Y.}~\bibnamefont {Ren}}, \bibinfo {author} {\bibfnamefont {H.~T.}\
  \bibnamefont {Stokes}}, \ and\ \bibinfo {author} {\bibfnamefont {T.~T.~M.}\
  \bibnamefont {Palstra}},\ }\href@noop {} { (\bibinfo {year}
  {2007}{\natexlab{a}})},\ \Eprint {http://arxiv.org/abs/cond-mat/0504546}
  {arXiv:cond-mat/0504546} \BibitemShut {NoStop}%
\bibitem [{\citenamefont {N$\acute{\textrm{e}}$nert}\ \emph
  {et~al.}(2007){\natexlab{b}}\citenamefont {N$\acute{\textrm{e}}$nert},
  \citenamefont {Pollet}, \citenamefont {Marinel}, \citenamefont {Blake},
  \citenamefont {Meetsma},\ and\ \citenamefont {Palstra}}]{Nenert2007}%
  \BibitemOpen
  \bibfield  {author} {\bibinfo {author} {\bibfnamefont {G.}~\bibnamefont
  {N$\acute{\textrm{e}}$nert}}, \bibinfo {author} {\bibfnamefont
  {M.}~\bibnamefont {Pollet}}, \bibinfo {author} {\bibfnamefont
  {S.}~\bibnamefont {Marinel}}, \bibinfo {author} {\bibfnamefont {G.~R.}\
  \bibnamefont {Blake}}, \bibinfo {author} {\bibfnamefont {A.}~\bibnamefont
  {Meetsma}}, \ and\ \bibinfo {author} {\bibfnamefont {T.~T.~M.}\ \bibnamefont
  {Palstra}},\ }\href@noop {} {\bibfield  {journal} {\bibinfo  {journal}
  {Journal of Physics-Condensed Matter},\ }\textbf {\bibinfo {volume} {19}},\
  \bibinfo {pages} {466212} (\bibinfo {year} {2007}{\natexlab{b}})}\BibitemShut
  {NoStop}%
\bibitem [{\citenamefont {Jeong}\ \emph {et~al.}(2007)\citenamefont {Jeong},
  \citenamefont {Hur},\ and\ \citenamefont {Proffen}}]{Jeong2007}%
  \BibitemOpen
  \bibfield  {author} {\bibinfo {author} {\bibfnamefont {I.~K.}\ \bibnamefont
  {Jeong}}, \bibinfo {author} {\bibfnamefont {N.}~\bibnamefont {Hur}}, \ and\
  \bibinfo {author} {\bibfnamefont {T.}~\bibnamefont {Proffen}},\ }\href@noop
  {} {\bibfield  {journal} {\bibinfo  {journal} {Journal of Applied
  Crystallography},\ }\textbf {\bibinfo {volume} {40}},\ \bibinfo {pages} {730}
  (\bibinfo {year} {2007})}\BibitemShut {NoStop}%
\bibitem [{\citenamefont {Kim}\ \emph {et~al.}(2009)\citenamefont {Kim},
  \citenamefont {Cho}, \citenamefont {Koo}, \citenamefont {Hong},\ and\
  \citenamefont {Shin}}]{Kim2009}%
  \BibitemOpen
  \bibfield  {author} {\bibinfo {author} {\bibfnamefont {J.}~\bibnamefont
  {Kim}}, \bibinfo {author} {\bibfnamefont {K.~C.}\ \bibnamefont {Cho}},
  \bibinfo {author} {\bibfnamefont {Y.~M.}\ \bibnamefont {Koo}}, \bibinfo
  {author} {\bibfnamefont {K.~P.}\ \bibnamefont {Hong}}, \ and\ \bibinfo
  {author} {\bibfnamefont {N.}~\bibnamefont {Shin}},\ }\href@noop {} {\bibfield
   {journal} {\bibinfo  {journal} {Applied Physics Letters},\ }\textbf
  {\bibinfo {volume} {95}},\ \bibinfo {pages} {132901} (\bibinfo {year}
  {2009})}\BibitemShut {NoStop}%
\bibitem [{\citenamefont {Lonkai}\ \emph {et~al.}(2004)\citenamefont {Lonkai},
  \citenamefont {Tomuta}, \citenamefont {Amann}, \citenamefont {Ihringer},
  \citenamefont {Hendrikx}, \citenamefont {{T\"obbens}},\ and\ \citenamefont
  {Mydosh}}]{Lonkai2004}%
  \BibitemOpen
  \bibfield  {author} {\bibinfo {author} {\bibfnamefont {T.}~\bibnamefont
  {Lonkai}}, \bibinfo {author} {\bibfnamefont {D.~G.}\ \bibnamefont {Tomuta}},
  \bibinfo {author} {\bibfnamefont {U.}~\bibnamefont {Amann}}, \bibinfo
  {author} {\bibfnamefont {J.}~\bibnamefont {Ihringer}}, \bibinfo {author}
  {\bibfnamefont {R.~W.~A.}\ \bibnamefont {Hendrikx}}, \bibinfo {author}
  {\bibfnamefont {D.~M.}\ \bibnamefont {{T\"obbens}}}, \ and\ \bibinfo {author}
  {\bibfnamefont {J.~A.}\ \bibnamefont {Mydosh}},\ }\href@noop {} {\bibfield
  {journal} {\bibinfo  {journal} {Phys. Rev. B},\ }\textbf {\bibinfo {volume}
  {69}},\ \bibinfo {pages} {134108} (\bibinfo {year} {2004})}\BibitemShut
  {NoStop}%
\bibitem [{\citenamefont {Abrahams}(2009)}]{Abrahams2009}%
  \BibitemOpen
  \bibfield  {author} {\bibinfo {author} {\bibfnamefont {S.~C.}\ \bibnamefont
  {Abrahams}},\ }\href@noop {} {\bibfield  {journal} {\bibinfo  {journal} {Acta
  Crystallographica Section B-structural Science},\ }\textbf {\bibinfo {volume}
  {65}},\ \bibinfo {pages} {450} (\bibinfo {year} {2009})}\BibitemShut
  {NoStop}%
\bibitem [{\citenamefont {Ibberson}\ \emph {et~al.}(1992)\citenamefont
  {Ibberson}, \citenamefont {David},\ and\ \citenamefont
  {Knight}}]{Ibberson1992}%
  \BibitemOpen
  \bibfield  {author} {\bibinfo {author} {\bibfnamefont {R.~M.}\ \bibnamefont
  {Ibberson}}, \bibinfo {author} {\bibfnamefont {W.~I.~F.}\ \bibnamefont
  {David}}, \ and\ \bibinfo {author} {\bibfnamefont {K.~S.}\ \bibnamefont
  {Knight}},\ }\href@noop {} {\emph {\bibinfo {title} {The High Resolution
  Powder Diffractometer (HRPD) at ISIS-a User Guide}}},\ \bibinfo {type}
  {Rutherford Appleton Laboratory Report}\ \bibinfo {number} {RAL-92–031}\
  (\bibinfo {year} {1992})\BibitemShut {NoStop}%
\bibitem [{\citenamefont {Ibberson}(2009)}]{Ibberson200947}%
  \BibitemOpen
  \bibfield  {author} {\bibinfo {author} {\bibfnamefont {R.~M.}\ \bibnamefont
  {Ibberson}},\ }\href@noop {} {\bibfield  {journal} {\bibinfo  {journal}
  {Nuclear Instruments and Methods in Physics Research Section A: Accelerators,
  Spectrometers, Detectors and Associated Equipment},\ }\textbf {\bibinfo
  {volume} {600}},\ \bibinfo {pages} {47 } (\bibinfo {year}
  {2009})}\BibitemShut {NoStop}%
\bibitem [{\citenamefont {Larson}\ and\ \citenamefont
  {Von~Dreele}(2000)}]{gsasref}%
  \BibitemOpen
  \bibfield  {author} {\bibinfo {author} {\bibfnamefont {A.~C.}\ \bibnamefont
  {Larson}}\ and\ \bibinfo {author} {\bibfnamefont {R.~B.}\ \bibnamefont
  {Von~Dreele}},\ }\href@noop {} {\emph {\bibinfo {title} {General Structure
  Analysis System (GSAS)}}},\ \bibinfo {type} {Los Alamos National Laboratory
  Report}\ \bibinfo {number} {LAUR 86-748}\ (\bibinfo {year}
  {2000})\BibitemShut {NoStop}%
\bibitem [{\citenamefont {Orobengoa}\ \emph {et~al.}(2009)\citenamefont
  {Orobengoa}, \citenamefont {Capillas}, \citenamefont {Aroyo},\ and\
  \citenamefont {Perez-Mato}}]{Orobengoa2009}%
  \BibitemOpen
  \bibfield  {author} {\bibinfo {author} {\bibfnamefont {D.}~\bibnamefont
  {Orobengoa}}, \bibinfo {author} {\bibfnamefont {C.}~\bibnamefont {Capillas}},
  \bibinfo {author} {\bibfnamefont {M.~I.}\ \bibnamefont {Aroyo}}, \ and\
  \bibinfo {author} {\bibfnamefont {J.~M.}\ \bibnamefont {Perez-Mato}},\
  }\href@noop {} {\bibfield  {journal} {\bibinfo  {journal} {Journal of Applied
  Crystallography},\ }\textbf {\bibinfo {volume} {42}},\ \bibinfo {pages} {820}
  (\bibinfo {year} {2009})}\BibitemShut {NoStop}%
\bibitem [{\citenamefont {Perez-Mato}\ \emph {et~al.}(2010)\citenamefont
  {Perez-Mato}, \citenamefont {Orobengoa},\ and\ \citenamefont
  {Aroyo}}]{Perez-Mato2010}%
  \BibitemOpen
  \bibfield  {author} {\bibinfo {author} {\bibfnamefont {J.~M.}\ \bibnamefont
  {Perez-Mato}}, \bibinfo {author} {\bibfnamefont {D.}~\bibnamefont
  {Orobengoa}}, \ and\ \bibinfo {author} {\bibfnamefont {M.~I.}\ \bibnamefont
  {Aroyo}},\ }\href@noop {} {\bibfield  {journal} {\bibinfo  {journal} {Acta
  Crystallographica Section A},\ }\textbf {\bibinfo {volume} {66}},\ \bibinfo
  {pages} {558} (\bibinfo {year} {2010})}\BibitemShut {NoStop}%
\bibitem [{\citenamefont {Campbell}\ \emph {et~al.}(2006)\citenamefont
  {Campbell}, \citenamefont {Stokes}, \citenamefont {Tanner},\ and\
  \citenamefont {Hatch}}]{Campbell2006}%
  \BibitemOpen
  \bibfield  {author} {\bibinfo {author} {\bibfnamefont {B.~J.}\ \bibnamefont
  {Campbell}}, \bibinfo {author} {\bibfnamefont {H.~T.}\ \bibnamefont
  {Stokes}}, \bibinfo {author} {\bibfnamefont {D.~E.}\ \bibnamefont {Tanner}},
  \ and\ \bibinfo {author} {\bibfnamefont {D.~M.}\ \bibnamefont {Hatch}},\
  }\href@noop {} {\bibfield  {journal} {\bibinfo  {journal} {Journal of Applied
  Crystallography},\ }\textbf {\bibinfo {volume} {39}},\ \bibinfo {pages} {607}
  (\bibinfo {year} {2006})}\BibitemShut {NoStop}%
\end{thebibliography}
\end{document}